\documentstyle[aps,pre,epsfig,twocolumn]{revtex}
\def\bea{\begin{eqnarray}}
\def\eea{\end{eqnarray}}
\def\beq{\begin{equation}}
\def\eeq{\end{equation}}

\begin{document}
\wideabs{
\title{Scaling of random spreading in small world networks}
\author{Jani Lahtinen$^1$, J\'anos Kert\'esz$^{1,2}$ and Kimmo Kaski $^1$}
\date{\today}
\maketitle
\begin{center}
{\it $^1$Laboratory of Computational Engineering, and 
Research Centre for Computational Science and Engineering,   
Helsinki University of Technology, P.O.Box 9400, FIN-02015 HUT, Finland}
\end{center}
\begin{center}
{\it $^2$Department of Theoretical Physics, Budapest University of
Technology, Budafoki \'ut 8, H-1111, Budapest, Hungary}
\end{center}
\maketitle
\begin{abstract}
In this study we have carried out computer simulations of random walks 
on Watts-Strogatz-type small world networks and measured the mean 
number of visited sites and the return probabilities. These quantities 
were found to obey scaling behavior with intuitively reasoned 
exponents as long as the probability $p$ of having a long range bond 
was sufficiently low. 

PACS numbers: 05.40.-a, 05.50.+q, 87.18.Sn
\end{abstract}
}
\bigskip

Recent interest in some disordered graphs was stimulated by a series
of discoveries related to the so called small world properties of
certain networks including social nets \cite{soc}, document retrieval 
in the www \cite{www}, the internet \cite{f3}, scientific cooperation 
relations \cite{redner-bar} etc. (for reviews see \cite {reviews}). 
Here 'small world' is used in the sense, that arbitrarily selected 
pairs of nodes can be reached from each other within few steps on the 
average, in spite of the relatively low number of links present in the 
system. 

Many of the earlier network studies have dealt with the Watts-Strogatz 
small world network (WSSWN) model, which we will also study. We define 
it as a one-dimensional ring consisting of $N$ nodes with $k$-neighbor 
interaction where, in addition, $N\cdot p$ new links are introduced 
between arbitrarily chosen, not yet connected nodes. Though the small 
world property is closely related to the dynamic process of information 
spreading in the network, relatively few papers have been devoted to 
this aspect \cite{WS,NW,dyn,JB}.

In the recent paper Jasch and Blumen \cite {JB} studied the target 
problem on a WSSWN model, i.e., the decay of the number of randomly 
distributed target sites on the graph where annihilating random walkers 
move around. There it turned out that this decay process is closely 
related to the mean number of distinct visited sites $S(n)$, where $n$ 
is the number of steps done by the walkers. In addition they observed 
the expected crossover in $S(n)$ from the $\sqrt{n}$ behavior that is 
characteristic for the one-dimensional case to $S(n)\propto n$ 
describing the high dimensional or random graph situation. $S(n)$ 
as a function of $p$ and $n$ has a scaling form:
$$S(n)=n^{1/2}f(np^\alpha) \eqno(1)$$ 
where $f$ is a universal scaling function with the following properties:
$$ f(x) \propto \left\{ \begin{array} {r@{\quad \mbox{for}\quad}l}
\mbox{const} & x \ll 1 \\ \sqrt{x} & x \gg 1 
\end{array} \right. \eqno(2)$$
Intuitively it is expected that $\alpha = 2$ since in the system 
there exists a basic length scale $\xi \propto 1/p$, characteristic of 
the average distance between nodes having long range links \cite{NW} 
for which the walker needs $n_\xi \propto \xi ^2$ steps to sweep
through. Thus the argument of the scaling function $f$ in Eq. (1) should
be $n/n_\xi$. In spite of this strong argument Jasch and Blumen \cite{JB} 
found in their numerical simulations of WSSWN's a value $\alpha = 1.85$. 
In the simulations they had chosen $N=50 000$ by taking an average over 
500 random walkers for each of the 100 WSSWN and they varied $p$ in the 
interval $0.01 \le p \le 0.1$. The main aim of the present note is to 
resolve the discrepancy in the value of $\alpha$.

It is first noted that Eq. (1), as is usual in scaling theory, is
valid only asymptotically and in this case the scaling limit is $N
\to \infty, n \to \infty$ and $p \to 0$. The scaling regime can be
estimated from the variation of the mean vertex distance $\ell$ as
a function of $p$ \cite{NW}: It turns out that $\ell k /N$ is a
scaling function with the argument $x = p\cdot k\cdot N$ and of
sigmoidal shape.  This curve suggests that one can not expect good
scaling for the above mentioned crossover, if $p\cdot k\cdot N \gg
100$.  Therefore, it seems likely that in \cite {JB} the
investigated values of $p$ were not small enough to assure the
proper scaling behavior ($p_{\rm min}\cdot k\cdot N = 1000$ in
\cite{JB}).

For this reason we have carried out our simulations with considerably
smaller values of $p$. In order to do so, we had to increase the
system size as well. Hence we have studied WSSWN's with $k=2$,
$N=10^5$ and for each value of $p=10^{-4}, 10^{-3.5}, 10^{-3},
10^{-2.5}$ we did $100$ realizations and had $100$ random walkers per
realization ($p_{\rm min}\cdot k\cdot N = 20$). In these simulations
we measured the average number of distinct visited sites $S(n)$ as
functions of $n$ and $p$, as depicted in Fig 1. In this plot it is seen 
that for the two largest values of $p$ saturation of $S(n)$ has set in.

\begin{figure}
  \centerline{\epsfxsize=8cm \epsfbox{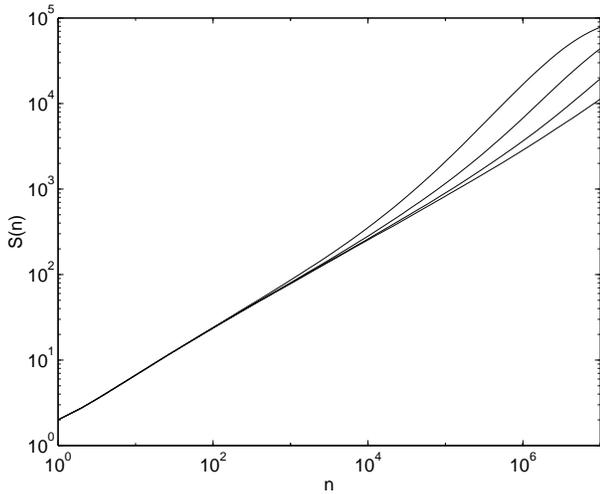}} \caption{Raw data
  for the average number of distinct sites visited $S$ as a function
  of time steps $n$ and the probability $p=10^{-4}, 10^{-3.5},
  10^{-3}, 10^{-2.5}$ increasing from bottom to top. For
  large $p$ the saturation due to the finite size $N=10^5$ of the
  systems is visible.}
\end{figure}

\begin{figure}
  \centerline{\epsfxsize=8cm \epsfbox{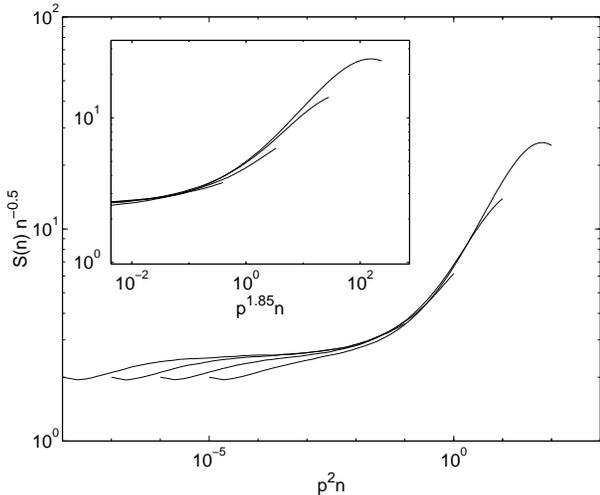}}
  \caption{Scaling plot of the data of Fig. 1 with $\alpha=2$. The
    inset shows the scaling plot with the exponent of 
    [9].}
\end{figure}

Fig. 2 shows a scaling plot of the results on $S(n)$ where
$S(n)/\sqrt{n}$ is plotted vs.  $np^\alpha$. The scaling was found to
be optimal with the choise of $\alpha = 2$. For comparison, we have
also shown the same plot with $\alpha = 1.85$, the value found in \cite
{JB}. Our results clearly support the simple scaling picture discussed
above. 

We have also checked the scaling behavior for another quantity, namely 
the return probability $P_{00}(n)$. This is known to decay as $1/\sqrt{n}$ 
for the $p=0$ case while an exponential decay is expected for large $p$. 
A scaling form similar to Eq. (1) should be valid for $P_{00}$:
$$ P_{00}= n^{-1/2}\varphi(x) \eqno(3)$$ 
where $ \varphi(x)$ is a
rapidly decaying scaling function with the limit $\varphi =
$ const. for $x \ll 1$.  However, the argument of
$\varphi$ should be the same as in eq. (1-2), namely $x=p^\alpha n$.

\begin{figure}
  \centerline{\epsfxsize=8cm \epsfbox{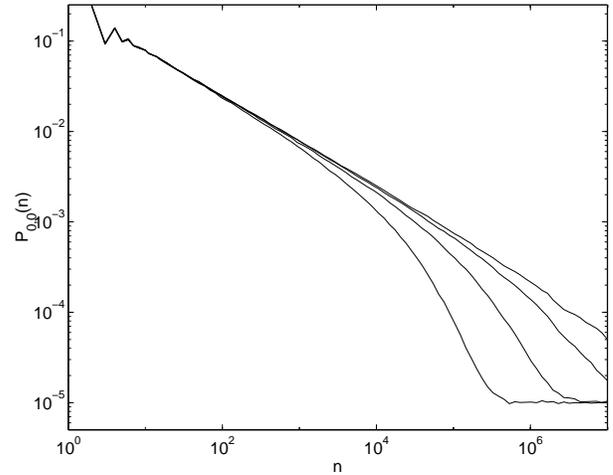}}
  \caption{Raw data for the return probability $P_{00}$. The $p$
  values are the same as for Fig. 1, now increasing from top to
  bottom.  The whole time interval was binned by 100 bins of equal
  sizes on the logarithmic scale.}
\end{figure}

\begin{figure}
  \centerline{\epsfxsize=8cm \epsfbox{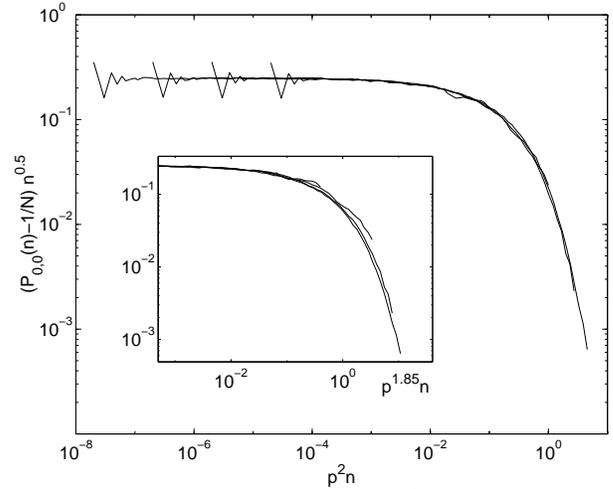}}
  \caption{Scaling plot of the data of Fig. 3 using $\alpha = 2$. For
  minimizing the finite size effects the asymptotic value $1/N=10^{-5}$ was
  subtracted from $P_{00}$. For comparison, the inset shows the
  scaling plot with the $\alpha$ of 
  [9].}
\end{figure}

Here in the interest of even higher accuracy we have used 10 times more 
runs for the averages. In order to minimize the effect due to the 
finite size of the samples, we have subtracted $1/N$, i.e., the 
$n\to \infty $ limit, from the measured values.  Fig. 3 shows the
data for the return probability $P_{00}$ and Fig. 4 the scaling plots. 
Again, we see that the scaling with the intuitively expected $\alpha = 2$ 
is superior to the one obtained by the value of \cite{JB}.

In conclusion we have shown that for sufficiently small probabilities
of long range links the proper scaling variable for the average number
of distinct sites visited and also for the return probability
is $np^2$, i.e., the natural exponent $\alpha = 2$ holds for the
Watts-Strogatz small world network.

This research was partially supported by OTKA T029985 and the Academy
of Finland's Centre of Excellence programme, No. 44897 (Research Centre 
for Computational Science and Engineering). JK thanks for the kind 
hospitality of Laboratory of Computational Engineering at HUT.


\end{document}